\documentclass[nofootinbib,showpacs,preprintnumbers,amsmath,amssymb,floatfix]
{revtex4}
\usepackage{epsf}

\begin{document}
\newcommand{\plb}[3]{{\it Phys.\ Lett.\ }{\bf B #1} (#2) #3}
\newcommand{\npb}[3]{{\it Nucl.\ Phys.\ }{\bf B #1} (#2) #3}
\renewcommand{\prd}[3]{{\it Phys.\ Rev.\ }{\bf D #1} (#2) #3}
\newcommand{\sjnp}[3]{{\it Sov.\ J.\ Nucl.\ Phys.\ }{\bf #1} (#2) #3}
\newcommand{\jetp}[3]{{\it Sov.\ Phys.\ JETP\/ }{\bf #1} (#2) #3}
\newcommand{\appol}[3]{{\it Acta Phys.\ Polon.\ }{\bf #1} (#2) #3}
\newcommand{\epjc}[3]{{\it Eur.\ Phys.\ J. }{\bf C #1} (#2) #3}
\newcommand{\ibid}[3]{{\it ibid.\ }{\bf #1} (#2) #3}
\newcommand{\zpc}[3]{{\it Z.\ Physik }{\bf C #1} (#2) #3}


\title{Comment on the recent COMPASS data on the spin
structure function $g_1$}

\vspace*{0.3 cm}

\author{B.I.~Ermolaev}
\affiliation{Ioffe Physico-Technical Institute, 194021
 St.Petersburg, Russia}
\author{M.~Greco}
\affiliation{Department of Physics and INFN, University Rome III,
Rome, Italy}
\author{S.I.~Troyan}
\affiliation{St.Petersburg Institute of Nuclear Physics, 188300
Gatchina, Russia}

\begin{abstract}
We examine the recent COMPASS data on the spin structure
function $g_1$ singlet. We show that it is rather difficult to use
the data in the present form in order to draw conclusions on the
initial parton densities. However, our tentative estimate is that
the data better agree with  positive rather than  negative initial
gluon densities.
\end{abstract}


\maketitle

\section{Introduction}

The COMPASS collaboration has recently presented new data,
Ref.~\cite{compass}, on the singlet
component of the spin structure function $g_1$. These data
were obtained from the measurements of the longitudinal spin
asymmetries in the scattering of muons off the $LiD$ target.
They found
that approximately
\begin{equation}\label{g1compass}
g_1 (x,Q^2) = 0,
 \end{equation}
with small errors, in a wide region of $x$ and at small $Q^2$.
More precisely, the kinematic
region  covered in Ref.~\cite{compass} is
\begin{equation}\label{compassreg}
G_{COMPASS}:\qquad 10^{-4} \lesssim x \lesssim 10^{-1};~~ 10^{-1}~GeV^2
\lesssim Q^2 \lesssim 1~GeV^2 .
\end{equation}
The fact that $g_1$ is zero in the wide region of $x$  at the
first sight  looks quite unexpected and even intriguing, and
clearly requires a theoretical explanation. The most-known
theoretical tool for describing $g_1$ is the Standard Approach
(SA), based on the DGLAP evolution equations, Ref.~\cite{dglap},
combined  with the standard fits, Refs.~\cite{fitsa}
-\cite{fitsb}, for the initial parton densities. However, the
values of $Q^2$ in the region $G_{COMPASS}$ are quite small and
therefore this region is beyond the reach of DGLAP. In
Refs.~\cite{egtsmallq,egtqcor} we have suggested an alternative
approach for describing $g_1$ at small $x$ and arbitrary $Q^2$.
Briefly, it combines the total resummation of the leading
logarithms of $x$ suggested in Refs.~\cite{egtns,egts} with the
shift
\begin{equation}\label{shiftq}
Q^2 \to \bar{Q^2} = Q^2 + \mu^2,
\end{equation}
 with $\mu \approx 5.5$ GeV for the singlet $g_1$.
 This shift automatically leads to an effective  change of $x$:
\begin{equation}\label{shiftx}
x \to \bar{x} = x +z
\end{equation}
where $z = \mu^2/w$,   $w \equiv 2pq$, and p, q are the proton and
virtual photon momenta respectively.  The shifts of $Q^2$ and $x$
in Eqs.~(\ref{shiftq},\ref{shiftx}) allowed us to express $g_1$
at small $x$ and arbitrary $Q^2$
in terms of $g_1^{LL}(x,Q^2)$ obtained in Refs.~\cite{egts}
by the total resummation of the leading logarithmic contributions in
the region of small $x$ and large $Q^2$: $g_1$ at small $x$ and
arbitrary $Q^2$ can be written as $g_1^{LL} (\bar{x},\bar{Q^2})$.
 Let us notice that
introducing a  shift, similarly to Eq.~(\ref{shiftq}),  has been a
common tool for describing the small- $Q^2$ kinematic region, see
e.g. Refs.~\cite{bad} and refs. therein. However, contrary to all
other approaches, we have introduced the shift
Eqs.~(\ref{shiftq},\ref{shiftx}) from the analysis of the  Feynman
graphs involved, and the value of $\mu$ is fixed from theoretical
considerations (see Refs.~\cite{egtsmallq, egtqcor} for detail).
In Ref.~\cite{egtsmallq} we have predicted that $g_1$ should not
depend on $x$ in the COMPASS kinematic region: indeed
Eq.~(\ref{shiftx}) shows that $g_1$ depends  on $z$ rather than
$x$ in the region $G_{COMPASS}$. The exact value of $g_1$ in this
region cannot be
 predicted
because it strongly depends on the interplay between the quark and
gluon contributions. Those contributions
involve the coefficient functions and the initial quark and gluon
densities $\delta q$ and
$\delta g$ which are unknown.
On the other hand, the very fact that $g_1 =0$ approximately, can
be used to estimate $\delta q$ and $\delta g$.  A straightforward
application of our results in Refs.~\cite{egtsmallq, egtqcor} to
the region $G_{COMPASS}$ might be misleading.The point is that the
approach of Refs.~\cite{egtsmallq, egtqcor} is valid in the
kinematic region of small $\bar{x}$, i.e. for $z \ll 1$, whereas
in the COMPASS experiments $30$ GeV$^2$ $\lesssim w \lesssim 270$
GeV$^2$ and therefore
\begin{equation}\label{zreg}
 1 \lesssim z \lesssim 0.1,
\end{equation}

In order to extend the approach of Refs.~\cite{egtsmallq,egtqcor}
to the region of Eq.~(\ref{zreg}) we suggest in the present paper
a simple interpolation expression for $g_1$ which combines the
approach of Refs.~\cite{egtsmallq, egtqcor} with accounting for the
non-logarithmic contributions to the coefficient functions
in the fixed orders in $\alpha_s$.

The present paper is organized as follows: in Sect.~II we remind
the expressions for $g_1$ obtained in Ref.~\cite{egtsmallq}; in
Sect.~III we generalize them to the region of Eq.~(\ref{zreg}),
adding non-logarithmic contributions to the coefficient functions
and anomalous dimensions.
Then, in Sect.~IV we apply this technique to the COMPASS data.
Sect.~V is for our concluding remarks.

\section{Expression for the singlet $g_1$ at small $\bar{x}$ and arbitrary $Q^2$
in the leading
logarithmic approximation}

Explicit expressions  for the singlet $g_1$ at small $\bar{x}$ and
arbitrary $Q^2$ in the Leading Logarithmic Approximation (LLA)
were obtained in Refs.~\cite{egtsmallq,egtqcor}. They account for
the total resummation of DL contributions and for the running
$\alpha_s$ effects. According to Ref.~\cite{egtsmallq}, the LLA
expression $(\equiv g_1^{LL})$ for the singlet $g_1$ at small $x$
and  arbitrary $Q^2$ is:
\begin{eqnarray}
\label{g1smallq} g_1^{LL}(\bar{x}, \bar{Q^2}) = &&\frac{<e^2_q>}{2}
\int_{- \imath \infty}^{\imath \infty} \frac{d \omega}{2 \pi
\imath} \Big(\frac{1}{z + x} \Big)^{\omega} \times \nonumber\\
\times && \Big[ \Big(C_q^{(+)}(\omega) \Big(\frac{Q^2 +
\mu^2}{\mu^2} \Big)^{\Omega_{(+)}} + C_q^{(-)}(\omega)
\Big(\frac{Q^2 +
\mu^2}{\mu^2} \Big)^{\Omega_{(-)}}\Big) \delta q(\omega) +\nonumber \\
&&+\Big(C_g^{(+)}(\omega)\Big(\frac{Q^2 + \mu^2}{\mu^2}
\Big)^{\Omega_{(+)}} + C_g^{(-)}(\omega)\Big(\frac{Q^2 +
\mu^2}{\mu^2} \Big)^{\Omega_{(-)}}\Big)\delta g(\omega) \Big]~.
\end{eqnarray}
The coefficient functions $C_q^{(\pm)}(\omega)$ and
$C_q^{(\pm)}(\omega)$ as well as the exponents
$\Omega_{(\pm)}(\omega)$ are expressed through the anomalous
dimensions $H_{ik}$ which account  for the total resummation of DL
contributions and the running $\alpha_s$ effects, as follows.
We remind here that in our approach we do not use the DGLAP
parametrization $\alpha_s = \alpha_s(Q^2)$. Instead, we use
the alternative we suggested in Ref.~\cite{egtalpha}. It
allows us to consider the region of really small $Q^2$ and
at the same time to be within the framework of the Perturbative QCD.

\subsection{Expressions for the exponents $\Omega_{(\pm)}$}
The explicit expressions for $\Omega_{(\pm)}$  are:
\begin{equation}
\label{omegapm} \Omega_{(\pm)} = \frac{1}{2} \big[ H_{qq} + H_{gg}
\pm \sqrt{R} \big]~,
\end{equation}
where
\begin{equation}\label{r}
R= (H_{qq} -  H_{gg})^2 + 4 H_{qg} H_{gq}
\end{equation}
and  $H_{qq},H_{gq},H_{qg},H_{gg}$ are the anomalous dimensions
calculated in LLA.
\subsection{Coefficient functions}
The expressions for the coefficient functions are also written in terms
of $H_{ik}$:
\begin{eqnarray}\label{cbk}
C_q^{(+)} = \frac{\omega(-X + \sqrt{R})}{2 T \sqrt{R}}~,\qquad
C_g^{(+)} = \frac{\omega H_{qg}}{T \sqrt{R}}~, \nonumber\\
C_q^{(-)} = \frac{\omega (X + \sqrt{R})}{2 T \sqrt{R}}~,\qquad
C_g^{(-)} = -\frac{\omega H_{qg}}{T \sqrt{R}} ~.
\end{eqnarray}
Here
\begin{equation}\label{t}
X = H_{gg}- H_{qq},~~ T=
\omega^2 - \omega (H_{gg} + H_{qq}) + (H_{gg}H_{qq} -
H_{gq}H_{qg})~
\end{equation}

\subsection{Anomalous dimensions}
Here we have:
\begin{eqnarray}\label{hik}
&& H_{qq} = \frac{1}{2} \Big[ \omega - Z + \frac{b_{qq} -
b_{gg}}{Z}\Big],\qquad H_{qg} = \frac{b_{qg}}{Z}~, \\ \nonumber &&
H_{gg} = \frac{1}{2} \Big[ \omega - Z - \frac{b_{qq} -
b_{gg}}{Z}\Big],\qquad H_{gq} =\frac{b_{gq}}{Z}~
\end{eqnarray}
where
\begin{equation}
\label{z}
 Z = \frac{1}{\sqrt{2}}\sqrt{(\omega^2 - 2(b_{qq} + b_{gg})) +
\sqrt{(\omega^2 - 2(b_{qq} + b_{gg}))^2 - 4 (b_{qq} - b_{gg})^2 -
16b_{gq} b_{qg} }}~,
\end{equation}
\begin{equation}
\label{bik} b_{ik} = a_{ik} + V_{ik}~,
\end{equation}
with the Born contributions $a_{ik}$ defined as follows:

\begin{equation}\label{aik}
a_{qq} = \frac{A(\omega) C_F}{2 \pi}~,\quad a_{qg} = \frac{
A'(\omega) C_F}{\pi}~,\quad a_{gq} = - \frac{n_f A'(\omega)}{2
\pi}~,\quad a_{gg} = \frac{4 N A(\omega)}{2 \pi}~,
\end{equation}
\begin{equation}
\label{vik} V_{ik} = \frac{m_{ik}}{\pi^2} D(\omega)~,
\end{equation}
\begin{equation}
\label{mik} m_{qq} = \frac{C_F}{2 N}~,\quad m_{gg} = - 2N^2~,\quad
m_{gq} = n_f \frac{N}{2}~,\quad m_{qg} = - N C_F~,
\end{equation}
and
\begin{equation}
\label{a} A(\omega) = \frac{1}{b} \Big[\frac{\eta}{\eta^2 + \pi^2}
- \int_0^{\infty} \frac{d \rho e^{-\omega \rho}}{(\rho + \eta)^2 +
\pi^2} \Big],
\end{equation}

\begin{equation}\label{aprime}
A'(\omega) = \frac{1}{b} \Big[ \frac{1}{\eta} - \int_{0}^{\infty}
 \frac{d \rho e^{- \omega \rho}}{(\rho + \eta)^2} \Big]~,
\end{equation}

\begin{equation}
\label{d} D(\omega) = \frac{1}{2 b^2} \int_{0}^{\infty} d \rho
e^{- \omega \rho} \ln \big( (\rho + \eta)/\eta \big) \Big[
\frac{\rho + \eta}{(\rho + \eta)^2 + \pi^2} + \frac{1}{\rho +
\eta}\Big],
\end{equation}
with $\eta=\ln(\mu^2/\Lambda_{QCD}^2)$ and $b=(33-2 n_f)/(12\pi)$.

\section{Expression for the singlet $g_1$ at arbitrary $\bar{x}$ and $Q^2$}

Our goal now is to obtain explicit expressions for the singlet
$g_1$ which could be valid at arbitrary $\bar{x}$ and $Q^2$. The
point is that the Eqs.~(\ref{cbk}) and (\ref{hik}) for the
coefficient functions and anomalous dimensions present the total
resummation of the leading logarithms of $\bar{x}$ but those contributions
are large when $\bar{x} \ll 1$ only. Alternatively, non-logarithmic
contributions can be
large at large $\bar{x} \lesssim 1$ and should be taken into account at
large $\bar{x}$. Such
terms are beyond the rich of our approach, so we cannot do the total
resummation of them. Instead, we can obtain them in
the  orders $\sim \alpha_s$ and $\sim \alpha^2_s$. Adding these
contributions to the expressions in Eqs.~(\ref{cbk}) and (\ref{hik}),
we arrive at new formulae for the
coefficient functions and anomalous dimensions, which are valid at arbitrary
$\bar{x}$.  In doing so, we
can use the DGLAP results for the anomalous dimensions and coefficient
functions. Let us demonstrate it in detail, using an example of the singlet
coefficient function $C_q$.
The dealing with the other coefficient function and anomalous dimensions
is quite similar. The NLO DGLAP singlet coefficient function
$C^{DGLAP}_q$ in the $\omega$ -space and at
integer $\omega = n$ is (see e.g. \cite{grsv})
\begin{equation}\label{cqdglap}
C^{DGLAP}_q = 1 + \frac{\alpha_s (Q^2)C_F}{2 \pi}\Big[-S_2(n) + (S_1)^2(n)
+ \Big(\frac{3}{2} - \frac{1}{n(n+1)}\Big)S_1(n) + \frac{1}{n^2} +
\frac{1}{2n} + \frac{1}{n+1} - \frac{9}{2}\Big]
\end{equation}
where we use the standard notations
\begin{equation}\label{sums}
S_1(n) = \sum_{k=1}^{n} 1/k~,\qquad S_2(n) = \sum_{k=1}^{n}
1/k^2~.
\end{equation}
The expression (\ref{cqdglap}) is obtained by direct calculation
of the Feynman graphs and is insensitive to the value of $Q^2$,
save the parametrization of $\alpha_s$.
So, we can borrow it, though after some appropriate changes: In the first place
it should be valid at arbitrary $\omega$; second, according to the
results of Ref.~\cite{egtalpha},  the coupling $\alpha_s (Q^2)$
should be changed to $A(\omega)$ defined in Eq.~(\ref{a}). The
analytic continuation of Eq.~(\ref{cqdglap}) to arbitrary $\omega$
is obtained through expressing the sums in Eq.~(\ref{sums}) in
terms of the polygamma $\psi$ -function and the Euler constant
$\textbf{C}$:
\begin{equation}\label{psi}
S_1(n) = \textbf{C} + n\psi (n-1),\qquad S_2(n-1) =
\frac{\pi^2}{6} + \psi'(n).
\end{equation}
After that we obtain an expression which we
address as  $C^{(1)}_q$ accounting for
both logarithmic and non-logarithmic contributions in the first loop.
Repeating the same procedure for the gluon coefficient function,
we obtain its first-loop value $C^{(1)}_g$. Apart from the
trivial replacement $S_1, S_2$ by $\psi (\omega)$ according to
Eq.~(\ref{psi}),  $C^{(1)}_q$
and $C^{(1)}_g$ differ from the NLO DGLAP coefficient functions
$C^{NLO~DGLAP}_{q,g}(\omega)$ by the treatment of $\alpha_s$:
\begin{equation}\label{cnlo}
C^{(1)}_q (\omega) = C^{NLO~DGLAP}_q(\omega)|_{\alpha_s \to
A}~,\qquad C^{(1)}_g (\omega) = C^{NLO~DGLAP}_g(\omega)|_{\alpha_s
\to A}~,
\end{equation}
with $A$ being defined in Eq.~(\ref{a}).
 The two-loop expressions $H^{(2)}_{ik}$
for the anomalous dimensions can be found quite similarly. They also
can be obtained from the NLO DGLAP anomalous dimensions
$\gamma^{NLO~DGLAP}_{ik}(\omega)$ with
expressing $S_1, S_2$ through $\psi(\omega)$ and replacing
$\alpha_s(Q^2)$ by $A(\omega)$:
\begin{equation}\label{hnlo}
\qquad H^{(2)}_{ik} (\omega) =
\gamma^{NLO~DGLAP}_{ik}(\omega)|_{\alpha_s \to A}~.
\end{equation}
Explicit expressions for the NLO DGLAP coefficient functions
and anomalous dimensions can be found e.g. in
Ref.~\cite{grsv}. Obviously, the replacement $\alpha_s(Q^2)$ by $A(\omega)$
in Eqs.~(\ref{cnlo},\ref{hnlo}) makes possible to use
$C^{(1)}_q, C^{(1)}_g$
and $H^{(2)}_{ik}$ at arbitrary $Q^2$ in contrast to the DGLAP
expressions for the coefficient functions and anomalous dimensions.
Combining $C^{(1)}_q, C^{(1)}_g$
and $H^{(2)}_{ik}$ with  Eqs.~(\ref{cbk},\ref{hik}),
we obtain the
interpolation formulae equally valid for small and large $\bar{x}$.
Indeed, the replacements $H_{ik}$ by $\widetilde{H}_{ik}$
and $C^{(\pm)}_{q,g}$ by $\widetilde{C}^{(\pm)}_{q,g}$ in
Eq.~(\ref{g1smallq}) allow to extend  the small-$\bar{x}$ formula
Eq.~(\ref{g1smallq}) to arbitrary $\bar{x}$. The new coefficient
functions $\widetilde{C}^{(\pm)}_{q,g}$ are defined as follows (the
superscripts $\pm$ are dropped here):
\begin{equation}\label{ctilde}
\widetilde{C}_q =C_q + C^{(1)}_q - \Delta C_q~,\qquad
\widetilde{C}_g =C_g + C^{(1)}_g - \Delta C_g
\end{equation}
where $C_{q,g}$ are defined in  Eq.~(\ref{cbk}), $\Delta
C_{q,g}$ are their perturbative first-loop expansions and
$C^{(1)}_{q},~C^{(1)}_{g}$ are given by Eq.~(\ref{cnlo}).
 The definitions  for new anomalous dimensions
$\widetilde{H}_{ik}$ look quite similar:
\begin{equation}\label{htilde}
\widetilde{H}_{ik} = H_{ik} + H^{(2)}_{ik} -\Delta H_{ik}
\end{equation}
where $H_{ik}$  are introduced in
Eq.~(\ref{hik}), $\Delta H_{ik}$ include the first and second terms of
their perturbative expansions  whereas
$H^{(2)}_{ik}$ are given by Eq.~(\ref{hnlo}). Now, introducing
$\widetilde{\Omega}_{(\pm)}$ according to Eq.~(\ref{omegapm}),  with
$\widetilde{H}_{ik}$ in place  of $H_{ik}$, we arrive at the expression
describing $g_1$ at arbitrary  $\bar{x}$ and $Q^2$:
\begin{eqnarray}
\label{g1arbxq} g_1(\bar{x}, \bar{Q^2}) = && \frac{<e^2_q>}{2} \int_{-
\imath \infty}^{\imath \infty} \frac{d \omega}{2 \pi \imath}
 \Big(\frac{1}{z + x}
\Big)^{\omega} \times \nonumber\\  \times && \Big[
\Big(\widetilde{C}_q^{(+)}(\omega) \Big(\frac{Q^2 + \mu^2}{\mu^2}
\Big)^{\widetilde{\Omega}_{(+)}} + \widetilde{C}_q^{(-)}(\omega)
\Big(\frac{Q^2 + \mu^2}{\mu^2}
\Big)^{\widetilde{\Omega}_{(-)}}\Big) \delta q(\omega) +
 \nonumber\\ &&+ \Big(\widetilde{C}_g^{(+)}(\omega)\Big(\frac{Q^2 + \mu^2}{\mu^2}
\Big)^{\widetilde{\Omega}_{(+)}} +
\widetilde{C}_g^{(-)}(\omega)\Big(\frac{Q^2 + \mu^2}{\mu^2}
\Big)^{\tilde{\Omega}_{(-)}}\Big)\delta g(\omega) \Big]~.
\end{eqnarray}

When $Q^2 \ll \mu^2$, Eq.~(\ref{g1arbxq}) can be expanded in the
series in $Q^2/\mu^2$:
\begin{equation}\label{g1combser}
 g_1(\bar{x},
\bar{Q^2}) \approx g_1(z) + (Q^2/\mu^2)
\frac{\partial g_1(\bar{x},\bar{Q^2})}{\partial Q^2/\mu^2} + O\big((Q^2/\mu^2)^2\big)
\end{equation}
where
\begin{equation}\label{g1comb}
g_1(z)= \frac{<e^2_q>}{2} \int_{- \imath
\infty}^{\imath \infty} \frac{d \omega}{2 \pi \imath}
 \Big(\frac{1}{z}
\Big)^{\omega} \Big[\widetilde{C}_q(\omega) \delta q +
\widetilde{C}_g(\omega) \delta g\Big].
\end{equation}
We have denoted here
 \begin{equation}\label{cqgcomb}
\widetilde{C}_q = \widetilde{C}^{(+)}_q + \widetilde{C}^{(-)}_q =
C_q + C^{(1)}_q - \Delta C_q, \qquad \widetilde{C}_g =
\widetilde{C}^{(+)}_g + \widetilde{C}^{(-)}_g = C_g + C^{(1)}_g
- \Delta C_g
\end{equation}
and
\begin{eqnarray}\label{cqcg}
&&C_q = C^{(+)}_q + C^{(-)}_q  = \frac{\omega (\omega - H_{gg})}
{\omega^2 - \omega (H_{gg} + H_{qq}) + H_{qq}H_{gg} -
H_{qg}H_{gq}}~, \quad \Delta C_q = 1 + \frac{a_{qq}}{\omega^2}~,
\nonumber \\ &&C_g = C^{(+)}_g + C^{(-)}_g  = \frac{\omega
H_{gq}}{\omega^2 - \omega (H_{gg} + H_{qq}) + H_{qq}H_{gg} -
H_{qg}H_{gq}}~, \quad \Delta C_g =  \frac{a_{gq}}{\omega^2}~.
\end{eqnarray}

\section{Implications
for  the recent COMPASS data.}

Eq.~(\ref{g1combser}) shows explicitly that $g_1$ practically does
not depend on $x$ in the region of Eq.~(\ref{compassreg}). It
perfectly agrees with  the flat $x$-dependence of $g_1$ observed
experimentally in Ref.~\cite{compass}.
Such a dependence means that $g_1$ in the COMPASS kinematic region
(\ref{compassreg}) does not depend on the conventional variables
$x$ and $Q^2$. On the contrary, Eq.~(\ref{g1combser}) predicts
that the $z$-dependence of $g_1$ is pretty far from being trivial.
Let us notice here that $z$ is inversely proportional to the
standard variable $\nu = w/(2M)$ measured in GeV, with $M=1$~GeV:
\begin{equation}\label{znu}
z = \Big(\frac{\mu^2}{2M}\Big) \frac{1}{\nu} \approx \frac{15}{\nu}~,
\end{equation}
so the region (\ref{zreg})  covered in the COMPASS experiment
corresponds to the $\nu$-region (in GeV)
\begin{equation}\label{nureg}
15 \lesssim \nu \lesssim 150.
\end{equation}
Obviously, a straightforward and unambiguous  application of our
description of $g_1$ to the COMPASS experiment could be obtained
just by fitting the COMPASS data on $g_1(z)$. Unfortunately, this
is impossible because the COMPASS collaboration has not studied
the $z$-dependence of $g_1$. Nevertheless, it is clear that
Eq.~(\ref{g1compass}) could be satisfied at any $z$ in the region
(\ref{zreg}) only if there exists a strong correlation between
$\delta q$ and $\delta g$ to compensate the difference between
$C_q$ and $C_g$ explicitly given in Eq.~(\ref{cqcg}). We think
that the chance for such a correlation is very tiny, though
strictly speaking this situation cannot be excluded. An
alternative interpretation  of the COMPASS result is to consider
Eq.~(\ref{g1compass}) as
\begin{equation}\label{g1av}
<g_1(z)> = 0.
\end{equation}
where $<g_1(z)>$ is the average value of $g_1$ observed by
COMPASS. Obviously, in order to match Eq.~(\ref{g1av}), $g_1(z)$
should acquire both positive and negative values in the region
(\ref{zreg}). For further investigations with Eq.~(\ref{g1av}) one
should choose appropriate fits for the initial parton densities
$\delta q (z)$ and $\delta g (z)$. Such fits are practically
absent in the literature. In Ref.~\cite{egtsmallq} we suggested to
approximate  $\delta q (z)$ and $\delta g (z)$ at small $z$ by
constants to get a rough  estimate. However, $z$ in the COMPASS
region (\ref{zreg}) is not small, so we prefer to use a DGLAP-like
set of  fits:
\begin{equation}\label{fitsb}
\delta q (z) = N_q z (1-z)^3 (1+3z),~~\delta g(z) = N_g (1-z)^4(1+3z).
\end{equation}

This set corresponds to the DGLAP-fits suggested in
Ref.~\cite{fitsa} but does not coincide with them. The difference
is in the power factors $z^a$ while the terms in the brackets in
Eq.~(\ref{fitsb}) and in Ref.~\cite{fitsa} coincide ( $x$ in
Ref.~\cite{fitsa} is replaced by $z$ in Eq.~(\ref{fitsb})). Indeed
 the fit for $\delta q$ in Ref.~\cite{fitsa} contains the singular
power factor $x^{-0.5}$ whereas the power factor  for $\delta g$
is $x^{0.5}$. In Ref.~\cite{egtinp} we have proved that the role
played by the singular terms $x^{-a}$ in the DGLAP fits is to
mimic the total resummation of $\ln^k(1/x)$~.   When the
resummation is taken into account, such factors ( namely the
factor $z^{-0.5}$ in $\delta g$) does not make sense any longer
and should be dropped. The same is obviously true when $x$ is
changed by $z$. So, extracting the singular factor $x^{-a}$ from
the fit in Ref.~\cite{fitsa}, we arrive at Eq.~(\ref{fitsb}).
Now it is easy to check that the fits (\ref{fitsb}) do not lead to
a flat $z$-dependence for  $g_1$ and cannot keep $g_1(z) =0$ in
the whole COMPASS region (\ref{compassreg}).

In more detail by substitution of  Eq.~(\ref{fitsb}) into
Eq.~(\ref{g1comb}) and performing the integration over $\omega$
numerically, with fixed and positive $N_q$ and varying the  values
of $N_g$, we plot our results in Fig. 1. By a close inspection of
the various configurations shown, we can easily conclude
 that these fits could be compatible  with Eq.~(\ref{g1av}) only
if $N_g > 0$ and $N_g > N_q$.

As the way of averaging $g_1$ over $z$ in the COMPASS data is
unknown, we can try another possibility, approximating
\begin{equation}\label{zav}
<g_1(z)> \approx g_1 (<z>) = 0,
\end{equation}
where $<z> = 0.25$ ~(i.e. $<\nu> \approx 60$ GeV) is the mean
value of $z$ from the region (\ref{zreg}). Then using
Eqs.~(\ref{g1comb},\ref{fitsb}), keeping  positive $N_q$ and
varying $N_g$. Figs.~1 suggest again that $N_g$ are positive and
$N_g > N_q$.

\section{Conclusion}

In the present paper we have considered in detail the recent
COMPASS data on $g_1$. These data first confirm our prediction in
Ref.\cite{egtqcor}  that $g_1$ at small $Q^2$ does not depend on
$Q^2$ and $x$. Instead, we predict that $g_1$ depends on the
invariant energy $w=2pq$ and the experimental investigation of
this dependence would allow to estimate the initial parton
densities. Unfortunately, this information is absent in the
present COMPASS data, so a reliable study of the initial parton
densities cannot be done. However, we have suggested  two possible
interpretations, Eqs.~(\ref{g1av}) and (\ref{zav}), of the COMPASS
result  Eq.~(\ref{g1compass}). Combining the LLA resummation with
the explicit first-loop values of the coefficient functions and using the DGLAP-like
parametrization (\ref{fitsb}) of the initial parton densities, we
conclude  that the data suggest  rather positive than negative
values of the initial gluon density. We remind that our analysis
is tentative. More quantitative conclusions can be drawn  only
after an accurate  experimental study of  the $z$-dependence of
$g_1$ has been performed.

\acknowledgments
 We are grateful to B.~Badelek, M.~Stolarski and
R.~Windmolders for their comments on the COMPASS experiments. The
work is partly supported by the Russian State Grant for Scientific
School RSGSS-5788.2006.2. Also we acknowledge partial support from
RTN European contracts MRTN-CT-2006-035482 ÒFLAVIAnetÓ and
MRTN-CT-2006-035505 ÒHeptoolsÓ.

\begin{figure}
\begin{center}
\begin{picture}(490,570)
\put(0,0){
\epsfbox{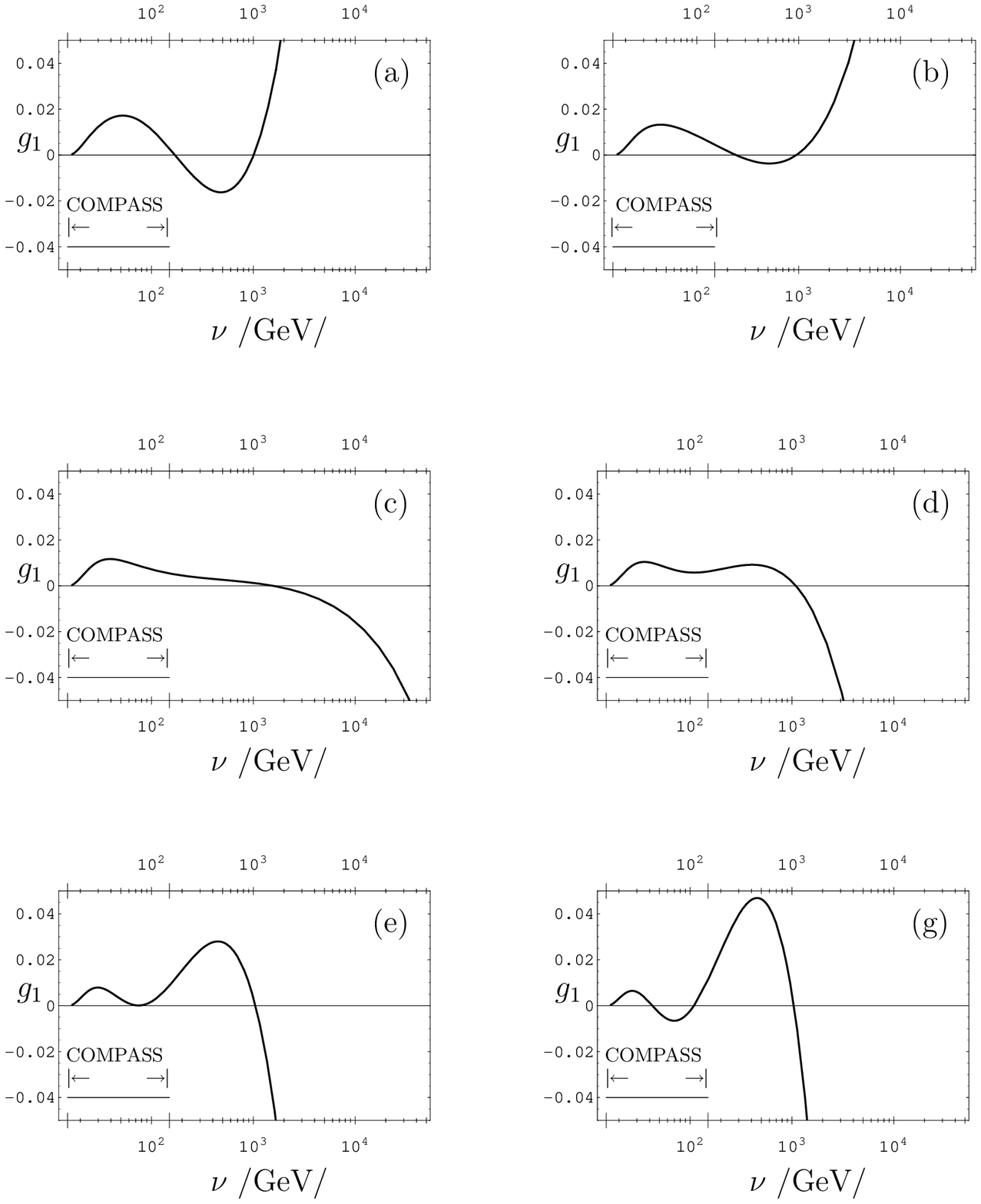}}
\end{picture}
\end{center}
\caption{The $\nu$ -dependence of $g_1(\nu)$, with $\delta q$,
$\delta g$ defined in Eq.~(\ref{fitsb}), for $N_q=0.5$ and
different values of $N_g$: (a) -1.5, (b) -0.5, (c) 0, (d) 0.5, (e)
2, (g) 3.5; the COMPASS $\nu$ -region corresponds to
Eq.~(\ref{nureg})~.} \label{fig1}
\end{figure}


\begin{thebibliography}{99}
\bibitem{compass} COMPASS Collaboration (E.S. Ageev et al),
{\it Spin asymmetry A1(d) and the spin-dependent structure
function g1(d) of the deuteron at low values of x and Q**2},\\
\plb {647}{2007}{330}.
\bibitem{dglap} G.~Altarelli and G.~Parisi, {\it Asymptotic Freedom in
Partonic Language}, \npb
{126}{1977}{298};\\
V.N.~Gribov and L.N.~Lipatov, {\it Deep inelastic e p scattering in
perturbation theory},\\ \sjnp {15}{1972}{438};\\
L.N.Lipatov, {\it The parton model and perturbation theory}, \sjnp
{20}{1974}{94};\\ Yu.L.~Dokshitzer, {\it  Calculation of the
Structure Functions for Deep Inelastic Scattering and e+ e-
Annihilation by Perturbation Theory in Quantum Chromodynamics (In
Russian)},\\ \jetp {46}{1977}{641}.
\bibitem{fitsa} G.~Altarelli, R.D.~Ball, S.~Forte and G.~Ridolfi,
{\it  Determination of the Bjorken sum and strong coupling from
polarized structure functions}, \npb {496}{1997}{337};\\
{\it  Theoretical analysis of polarized structure functions},
\appol {B29}{1998}{1145}.
\bibitem{fitsl} E.~Leader, A.V.~Sidorov and D.B.~Stamenov,
{\it  Longitudinal polarized parton densities updated}, \prd
{73}{2006}{034023}.
\bibitem{fitsb}J.~Blumlein and H.~Botcher, {\it QCD analysis of polarized
deep inelastic data and parton distributions}, \npb {636}{2002}{225};\\
M.~Hirai at al., {\it Determination of polarized parton
distribution functions and their uncertainties}, \prd
{69}{2004}{054021}.
\bibitem{egtsmallq} B.I.~Ermolaev, M.~Greco and S.I.~Troyan,
{\it Singlet structure function g(1) at small x and small Q**2},
\epjc {50}{2007}{823}.
\bibitem{egtqcor} B.I.~Ermolaev, M.~Greco and S.I.~Troyan,
{\it Perturbative power Q**2-corrections to the structure function
g(1)}, \epjc {51}{2007}{859}.
\bibitem{egtns} B.I.~Ermolaev, M.~Greco and S.I.~Troyan,
 {\it Intercepts of the nonsinglet structure functions},
 \npb {594}{2001}{71}; {\it QCD running coupling effects for the
 nonsinglet structure functions at small x}, \ibid {571}{2000}{137}.
\bibitem{egts} B.I.~Ermolaev, M.~Greco and S.I.~Troyan, {\it
Running coupling effects for the singlet structure function
g(1) at small x}, \plb {579}{2004}{321}.
\bibitem{bad} B.~Badelek and J.~Kwiecinski, {\it Analysis Of The
Electroproduction Structure Functions In The Low Q**2 Region
Combining The Vector Meson Dominance And The Parton Model With
Possible Scaling Violation}, \zpc {43}{1989}{251};\\ {\it Low
Q**2, low x region in electroproduction: An Overview}, \rmp
{68}{1996}{445};\\ {\it Unified description of the nonsinglet spin
dependent structure function g1 incorporating Altarelli-Parisi
evolution and the double logarithmic ln**2 (1/x) effects at low
x},\\ \plb {418}{1998}{229}.
\bibitem{grsv} M.~Gluck, E.~Reya, M.~Stratmann and
W.~Vogelsang, {\it Next-to-Leading Order Radiative Parton Model
Analysis of Polarized Deep Inelastic Lepton Nucleon Scattering},
\prd {63}{1996}{4775}.
\bibitem{egtinp} B.I.~Ermolaev, M.~Greco and S.I.~Troyan,
{\it Non-singlet structure functions: Combining the leading logarithms
resummation at small-x with DGLAP}, \plb {622}{2005}{93}.
\bibitem{egtalpha} B.I.~Ermolaev, M.~Greco and S.I.~Troyan,
{\it Treatment of the QCD coupling in high energy processes},
\plb {522}{2001}{57}.
\end{thebibliography}
\end{document}